\documentclass[preprint,12pt]{elsarticle}
\makeatletter
\def\ps@pprintTitle{%
 \let\@oddhead\@empty
 \let\@evenhead\@empty
 \def\@oddfoot{\centerline{\thepage}}%
 \let\@evenfoot\@oddfoot}
\makeatother

\usepackage{graphicx}
\usepackage{amssymb}
\usepackage{subfig}
\usepackage[margin=1.2in,footskip=0.25in]{geometry}
\usepackage{adjustbox}

\begin{document}

\begin{frontmatter}


\title{Visual Exploration and Knowledge Discovery from Biomedical Dark Data}


\author[1]{Shashwat Aggarwal}
\address[1]{University Of Delhi}
\ead{shashwata.co@nsit.net.in}
\author[2]{Ramesh Singh}
\address[2]{National Informatics Center}
\ead{rsingh@nic.in}

\begin{abstract}
Data visualization techniques proffer efficient means to organize and present data in graphically appealing formats, which not only speeds up the process of decision making and pattern recognition but also enables decision-makers to fully understand data insights and make informed decisions. Over time, with the rise in technological and computational resources, there has been an exponential increase in the world’s scientific knowledge. However, most of it lacks structure and cannot be easily categorized and imported into regular databases. This type of data is often termed as Dark Data. Data visualization techniques provide a promising solution to explore such data by allowing quick comprehension of information, the discovery of emerging trends, identification of relationships and patterns, etc. In this empirical research study, we use the rich corpus of PubMed comprising of more than 30 million citations from biomedical literature to visually explore and understand the underlying key-insights using various information visualization techniques. We employ a natural language processing based pipeline to discover knowledge out of the biomedical dark data. The pipeline comprises of different lexical analysis techniques like Topic Modeling to extract inherent topics and major focus areas, Network Graphs to study the relationships between various entities like scientific documents and journals, researchers, and, keywords and terms, etc. With this analytical research, we aim to proffer a potential solution to overcome the problem of analyzing overwhelming amounts of information and diminish the limitation of human cognition and perception in handling and examining such large volumes of data.
\end{abstract}


\end{frontmatter}


\section{Introduction}
\label{S:1}

In today’s data centralized world, the practice of data visualization has become an indispensable tool in numerous domains such as \textit{Research, Marketing, Journalism, Biology,} etc. Data visualization is the art of efficiently organizing and presenting data in a graphically appealing format. It speeds up the process of decision making and pattern recognition, thereby enabling decision makers to make informed decisions. With the rise in technology, the data has been exploding exponentially, and the world’s scientific knowledge is accessible with ease. There is an enormous amount of data available in the form of scientific articles, government reports, natural language, and images that in total contributes to around 80\% of overall data generated globally. However, most of the data lack structure and cannot be easily categorized and imported into regular databases. This type of data is often termed as Dark Data. Data visualization techniques proffer a potential solution to overcome the problem of handling and analyzing overwhelming amounts of such information. It enables the decision maker to look at data differently and more imaginatively. It promotes creative data exploration by allowing quick comprehension of information, the discovery of emerging trends, identification of relationships and patterns, etc.

In this empirical research, we visually explore and mine knowledge out of biomedical research documents present in the rich corpus of PubMed. Firstly, we use text summarization and visualization techniques like computation of raw term and document frequencies, streamline analysis for most frequent terms inside the corpus, word clouds, etc., to get a general overview of the PubMed database. We then utilize the MALLET library,  \citep{AuthorYear} to perform topic modeling to extract knowledge from the biomedical database by identifying and clustering the major topics inherent in the database. Finally, we use VOSviewer network viewer, \citep{Author2year2} to construct bibliometric networks for studying relationships between different entities like scientific documents and journals, researchers, and, keywords and terms.

\section{Visual Exploration of PubMed}
\label{S:2}

We use the rich corpus of the PubMed which “\textit{comprises of more than 30 million citations and abstracts for biomedical literature from MEDLINE, life science journals, and online books.}”  \citep{Author3year3}. It is an open source database developed and maintained by NCBI. In addition to free access to MEDLINE, PubMed also provides links to free full-text articles provided by PubMed Central and third-party websites, and other facilities such as clinical queries search filters, special query pages, etc. “\textit{PubMed is a key information resource in biological sciences and medicine primarily because of its wide diversity and manual curation. It comprises of an order of three billion bases of the human genome, rich meta-information (e.g., MeSH terms), detailed affiliation, etc., summing up to a total of 70GB database.}” \citep{Author4year4}. As of 1 August 2020, PubMed has more than 30 million records from 5500 journals, dating back to 1966, with the earliest publication available from the year 1809. PubMed supports efficient methods to search the database by using author names, journal names, keywords, and phrases, or any combination of these. It also enables users to download the fetched citations and abstracts for queried terms in various formats such as plain text form (both Summary and Abstract), XML form, PMID form, CSV form and MEDLINE form.

In Table~\ref{table:t1}, we present some basic statistics to describe the PubMed database – the size of the database, number of documents present inside the database, number of sentences and terms summed across all the documents present in the database. We also compute the number of distinct literals, entities, subjects and objects present within PubMed using a natural language processing (NLP) based pipeline: a) scraping of data in HTML/XML format (i.e., Parsing), b) striping it into plain text, and c) applying NLP pre-processing techniques such as sentence segmentation and word tokenization to compute the number of distinct literals, stemming and lemmatization for normalization of terms and phrases, POS (part of speech) tagging and Dependency Parsing to identify subjects and objects within each sentence, and finally, NER (named entity recognition) and Coreference Resolution to identify the entities present inside the database. 

\begin{table}[t]
\centering
\caption{Fundamental Statistics of PubMed}
\begin{tabular}{l|l|l|l}
\hline
\textbf{Size}         & 70 GB       & \textbf{\# Distinct Literals} & 1,842,783,647 \\ \hline
\textbf{\# Documents} & 359,324     & \textbf{\# Distinct Entities} & 412,593,720   \\ \hline
\textbf{\# Sentences} & 110 Million & \textbf{\# Distinct Subjects} & 412,593,720   \\ \hline
\textbf{\# Words}     & 2.7 Billion & \textbf{\# Distinct Objects}  & 436,101,294  \\
\hline
\end{tabular}
\label{table:t1}
\end{table}

In order to visually explore and analyse the biomedical research document present within PubMed, we firstly use Word Clouds, a visual representation of text data, typically used to depict the prominent words across the data with the prominence of words measured relative to their frequency counts, to summarize and get a general overview of the contents of PubMed documents. Word clouds are commonly used in the field of text mining and information retrieval for abstracting, visualizing, and comparing textual databases and have demonstrated to be useful in various research settings. We use the open source tool, Wordle  \citep{Author6year6} to create cloud visualization to summarize the contents of the PubMed corpora.

\begin{figure}[!htb]
\centering
\subfloat[ ][ ]{\includegraphics[scale=0.4]{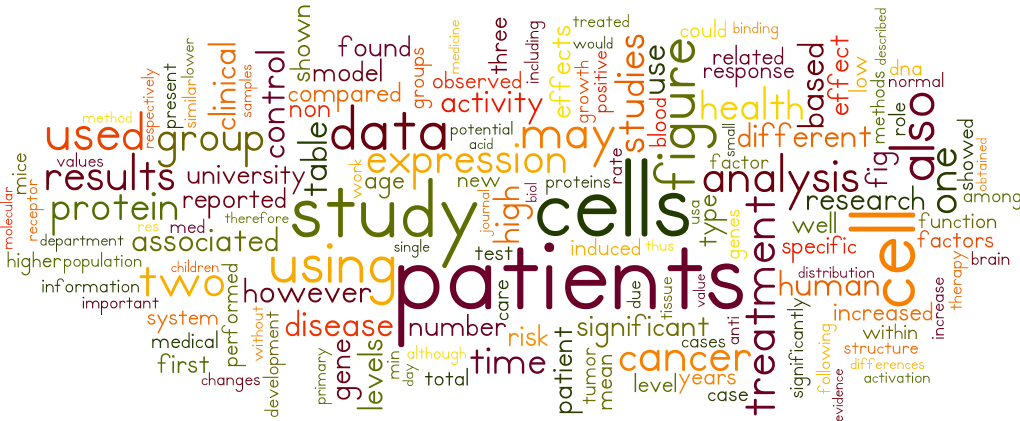}} \\
\subfloat[ ][ ]{\includegraphics[scale=0.4]{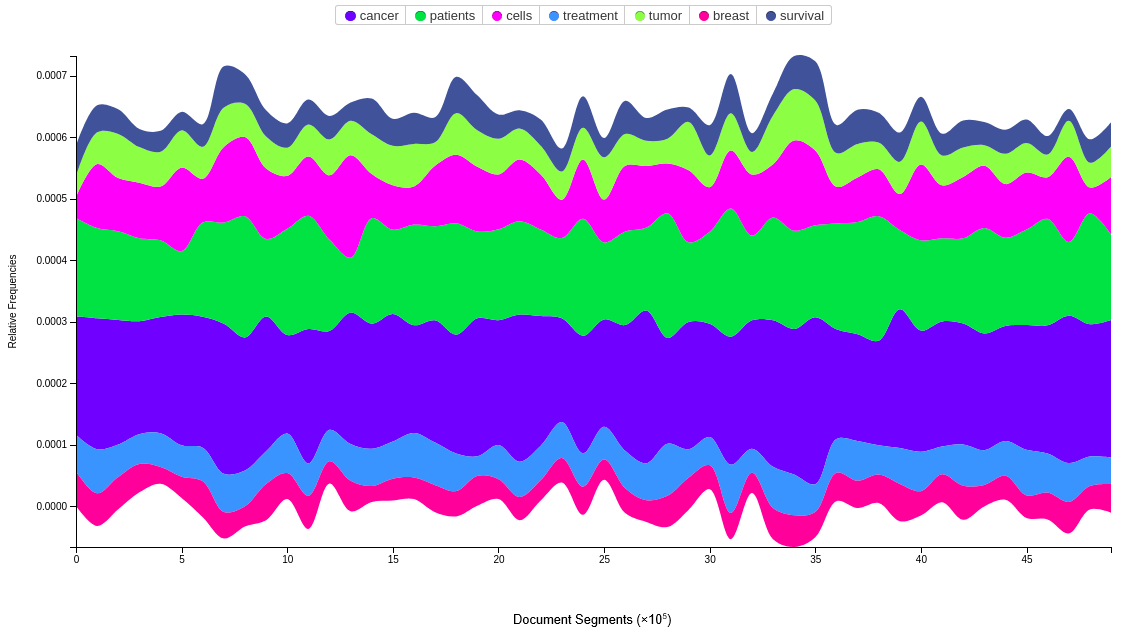}}
\caption{(a): Word Cloud of the PubMed Corpora. (b): Streamline Graph for top 7 most frequent words in PubMed.}
\label{fig:f1}
\end{figure}

Figure~\ref{fig:f1} presents the word cloud on the PubMed dataset. Words like \textit{patient, cells, data, cancer, gene} are more prominent signifying a substantial proportion of study related to cancer and genomics being conducted compared to other domains.  Furthermore, various words such as \textit{DNA, tumour, acid,} and \textit{receptors} highlight the other significant areas where research has been done or is going on. One interesting fact that can be observed from these clouds is the prominence of words, \textit{High Population} and \textit{Children} providing a high-level indication of the major disease cause and majority group affected by those diseases. Alongside the word cloud, in Figure~\ref{fig:f1} we also plot the streamline graph for the seven most frequent terms across the database depicting the variation of their relative frequency distribution across the set of documents.

Another widely used text representation technique which apart from considering the term frequencies, also encodes their importance inside a document, is $TF\times IDF$. The TF gives the frequency of the term within a particular document, and the IDF gives the inverse document frequency of a term, i.e., a measure of the importance of term across several documents. $TF\times IDF$, term frequency-inverse document frequency, is a statistical measure used to evaluate how important a word is to a document in a collection. The importance is directly proportional to the term frequency of the word in the document but is offset by the frequency of the word in the corpus.

\begin{figure}[!tb]
\begin{center}
\includegraphics[scale=0.28]{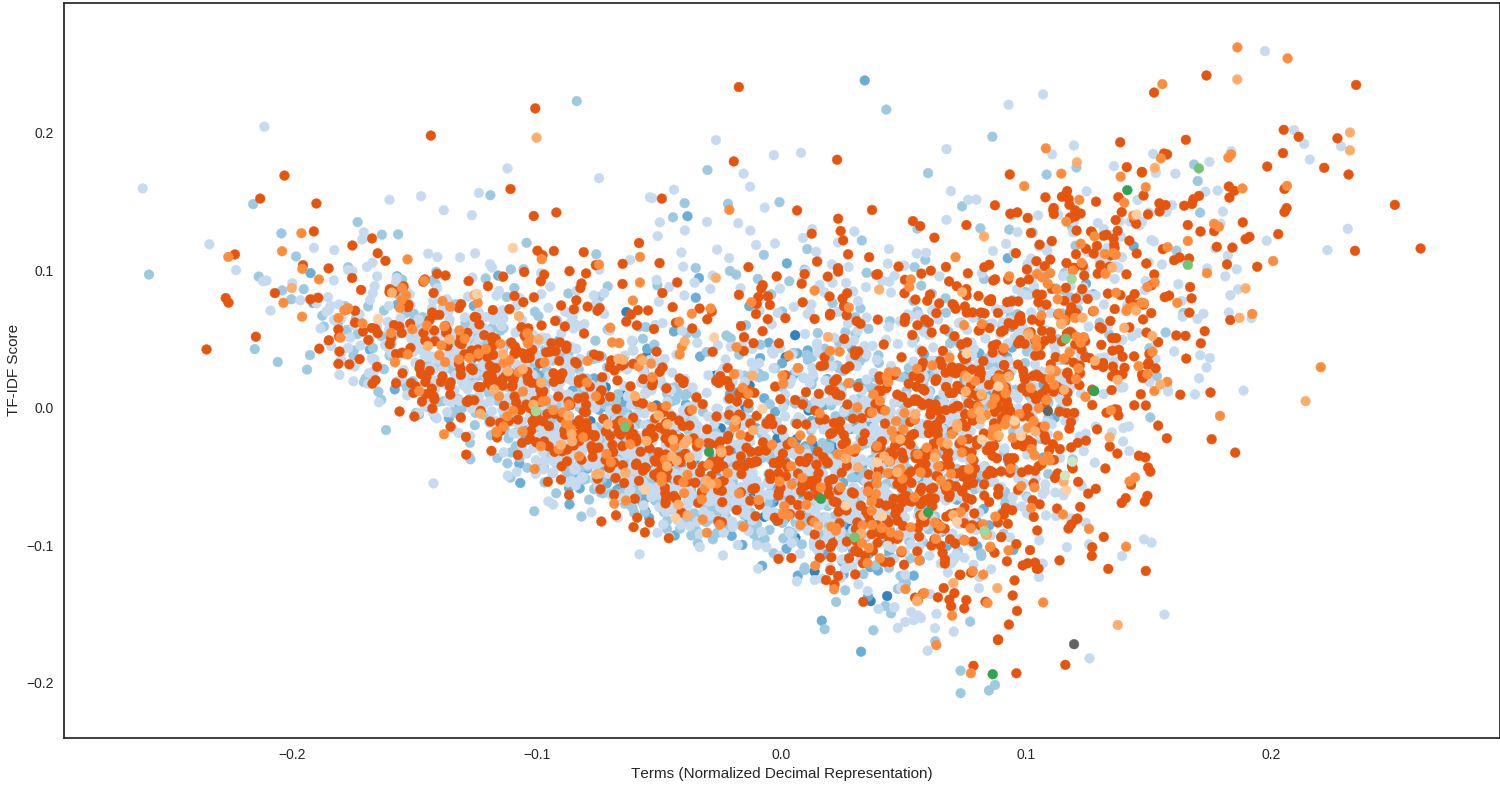}
\caption{$TF\times IDF$ Score Distribution Plot over 50,000 full-text articles retrieved from PubMed Central.  }
\label{fig:f2}
\end{center}
\end{figure}

\begin{figure}[!tb]
\centering
\subfloat[ ][ ]{\includegraphics[scale=0.25]{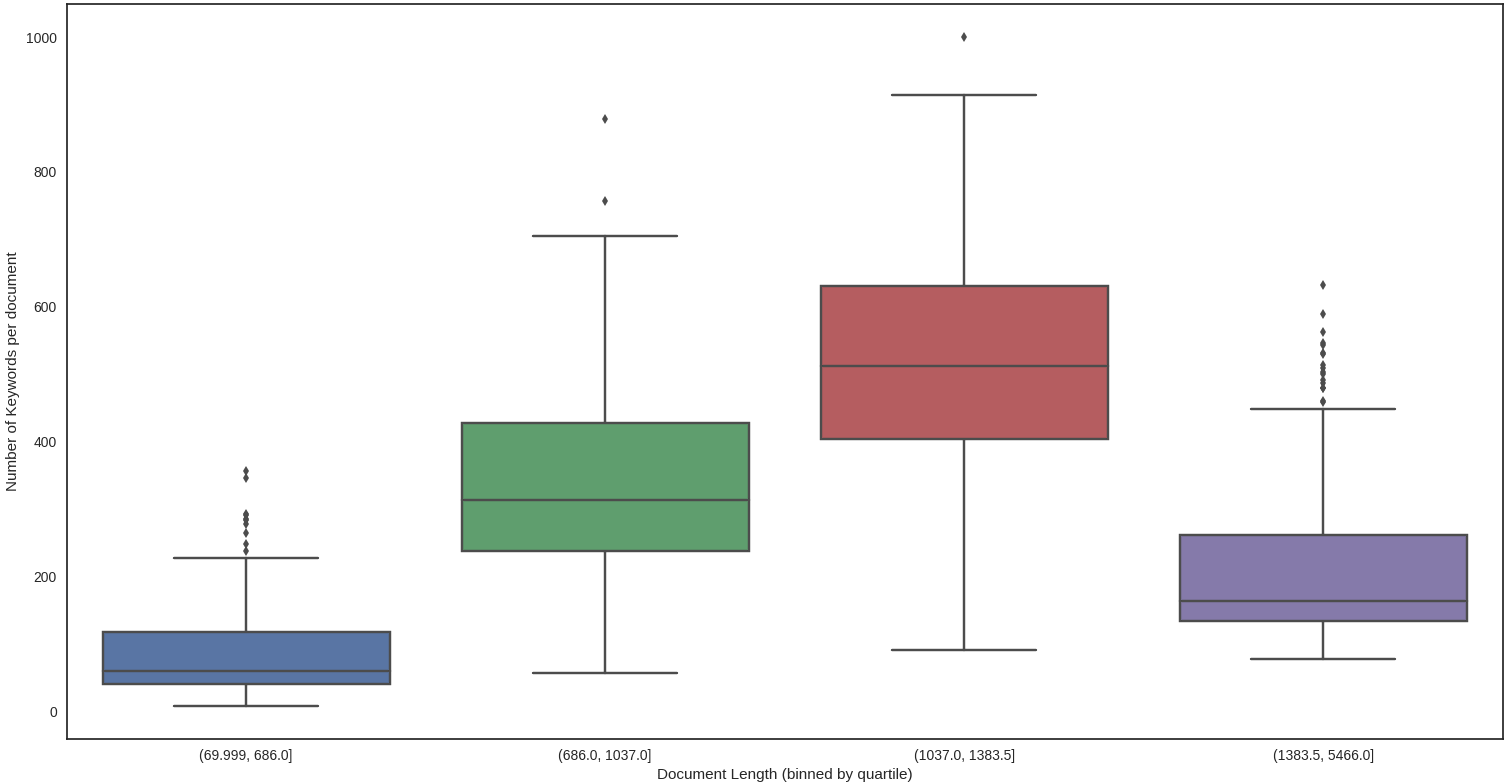}} \\
\subfloat[ ][ ]{\includegraphics[scale=0.25]{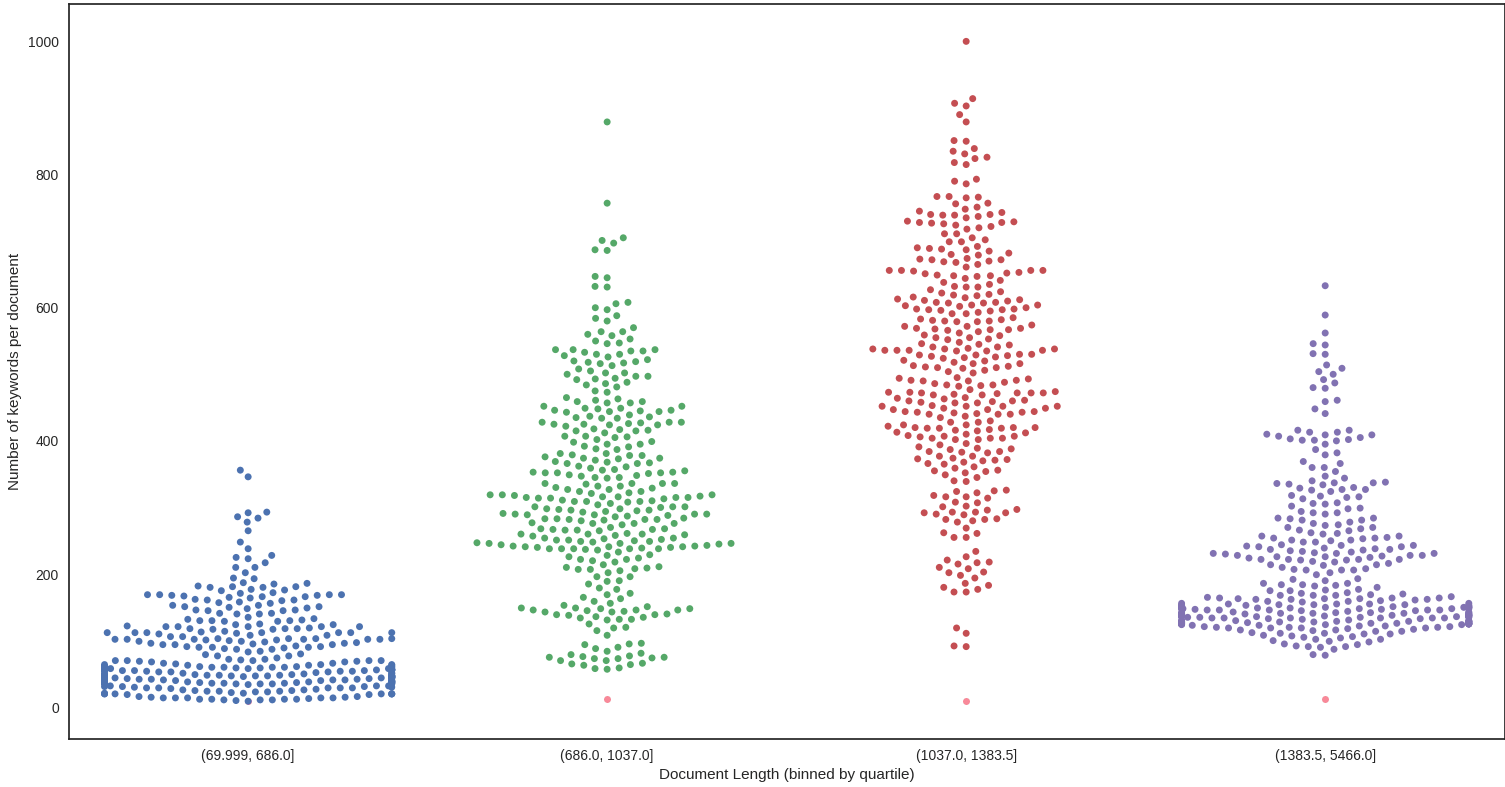}}
\caption{(a): Box-and-Whisker plot and (b): Swarm Plot, computed over 50,000 full text articles from PubMed Central showing the distribution of keywords across different document lengths (binned by quartile.)}
\label{fig:f3}
\end{figure}

In Figure~\ref{fig:f2}, we visualize the $TF\times IDF$ plot computed over 50,000 full text articles retrieved from PubMed Central. The x-axis represents the normalized decimal term representations while the y axis gives their corresponding $TF\times IDF$ scores. All the $TF\times IDF$ scores are calculated by calculating the term document matrix using Sklearn $TF\times IDF$ vectorizer \citep{scikit-learn}. The top 2 most significant components of the corresponding vector representation of the terms obtained from $TF \times IDF$ vectorizer are computed using PCA (principal component analysis) dimensionality reduction technique \citep{Jolliffe} and are visualized in Figure~\ref{fig:f2}. We extract a number of relevant words (A.K.A. keywords) in accordance to the $TF\times IDF$ scores computed earlier, with the number determined by a certain threshold score. We visualize the distribution of the number of keywords extracted via $TF\times IDF$ score for different document lengths. score for different document lengths. For space constraints and sparsity reasons, we binned the document lengths by quartile (i.e., the bins are not of equal range but contains the same number of documents – 25\% each). Figure~\ref{fig:f3} displays the box-and-whisker plot computed over 50,000 full text articles from PubMed Central showing the distribution of keywords across different document lengths (binned by quartile). Figure~\ref{fig:f3} also shows the swarm plot, which gives a better representation of the distribution of keywords, visualizing all observations along with the underlying distribution.

As it can be seen from the plots, the median value of number of keywords increase with the document length till the third quartile after which there is a drop in the median value, which is mainly because of  $TF\times IDF$ scoring and is intuitive since a given word is more likely to be found in a relatively longer document as compared to a shorter document but is not necessarily a keyword. We can also observe that the variation of number of keywords is less in first and the last bins as compared to the second and the third bin, indicating that documents which are either too short or long approximately contain a constant number of relevant words while moderate length documents have a high variability in their distribution of the number of keywords.

Although the raw term frequencies, word cloud visualizations and $TF\times IDF$ scores work quite well in practice (e.g., in summarization of general overview of the database), however, these techniques fail to capture the ordered relationship between terms and sentences. To understand the contextual relationship between the different terms we use DocuBurst. DocuBurst \citep{Author9year9} is an online document visualization tool used for creating interactive visual summaries of documents, exploring keywords to uncover document themes or topics, investigating intra-document word patterns, such as character relationships, comparing documents, etc. which takes advantage of the human-created structure in lexical databases. DocuBurst visualize nouns and phrases in a hierarchically structured manner centered around a root word which is selected either as the most prominent word in the database or as queried by the user. DocuBurst uses a pre-existing ontology, WordNet \citep{Author10year10}, to group words having related meanings together. It creates a radial, space-filling layout of hyponymy (IS-A relation) with interactive techniques of zoom, filter, and details-on-demand for the task of document visualization.

\begin{figure}[!htb]
\begin{center}
\includegraphics[scale=0.4]{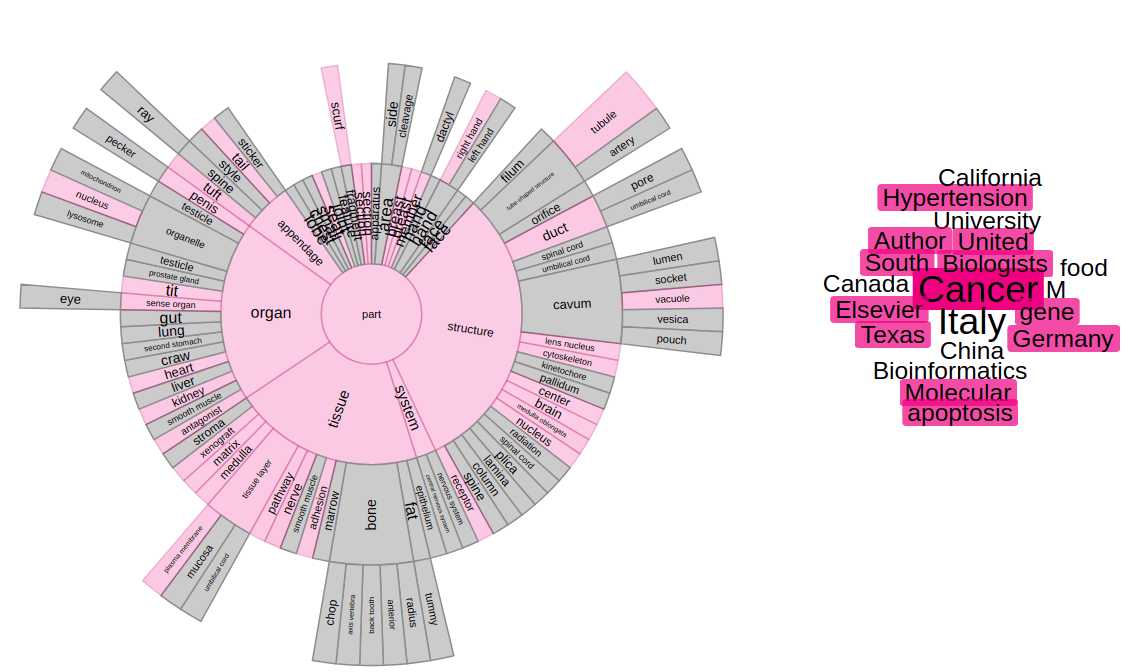}
\caption{DocuBurst plot on the PubMed database with \textit{part} as the root word. }
\label{fig:f4}
\end{center}
\end{figure}

We generate DocuBurst graphs over 50,000 full-text articles that we retrieved earlier from PubMed Central. We limit our database to this subset of PubMed to meet the software and memory requirements of the tool utilized. Alongside the DocuBurst graphs, both word score and word clouds for the selected word (shown in pink) and its co-occurring words are also displayed to summarize the content better. Figure ~\ref{fig:f4} shows the DocuBurst graph with part chosen as the root word. DocuBurst hierarchically structures the radially surrounding hyponyms such as organ, structure, tissue, and system around the root word. Each surrounding hyponym is further sub-structured with its related hyponyms, thereby forming chains of correlated keyword terms which reveal the coherent document themes present in the database. Along with the DocuBurst graph, Word Clouds depicting words having strong correlations with terms on the DocuBurst graph are also shown. From these word clouds, we can infer that words like \textit{Cancer} and \textit{Hypertension} are highly correlated with terms related to body parts. The word \textit{Cancer} can be seen to be highly connected with term \textit{tissues} and also with other terms like \textit{University, Research} and specific country names such as \textit{China, Germany} indicating significant work related to cancer research being carried out by Universities in these countries. Similarly, it can also be observed that terms like \textit{Hypertension} are highly correlated to terms highlighted in pink which are mostly related to \textit{mind} thus, indicating some of the body organs affected due to hypertension.

\begin{figure}[!htb]
\begin{center}
\includegraphics[scale=0.3]{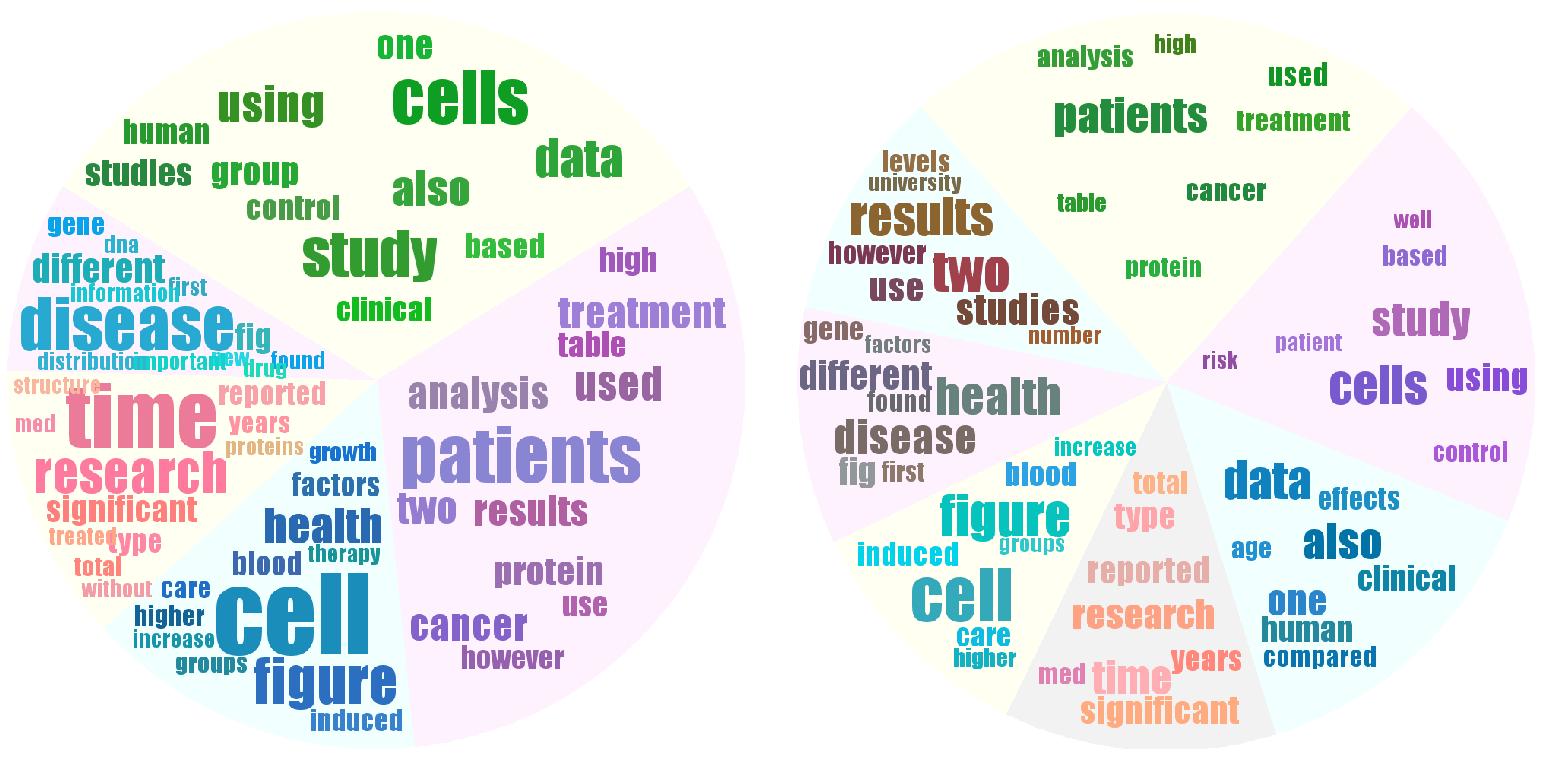}
\caption{Topic Cloud for Left: \textit{n=5} and Right: \textit{n=7} topics respectively. }
\label{fig:f5}
\end{center}
\end{figure}

\begin{table}[!t]
\caption{Top 10 topics along with their proportion in PubMed; Computed using MALLET.}
\vspace{2mm}
\begin{tabular}{c|c|p{10cm}}
\hline
\textbf{Topic Id} & \textbf{Proportion} & \multicolumn{1}{c}{ \textbf{Major Words in topic}}  \\
\hline 
1        & \textbf{0.26558}          & cancer clinical review studies therapeutic development disease treatment potential current molecular research provide including recent strategies data role approach diseases    \\
2        & 0.11401          & patients survival cancer analysis prognostic stage study factors group lymph significantly tumor clinical multivariate risk significant gastric metastasis ratio compared        \\
3        & 0.10156          & cells cell cancer expression apoptosis proliferation growth migration protein pathway effect lines study effects human signaling inhibition invasion tumor induced               \\
4        & 0.09279          & cancer risk women breast study years age screening mortality association incidence factors data population higher increased men diagnosis health rates                           \\
5        & 0.08076          & cancer care patients health quality study life treatment patient oncology medical information studies data intervention research systematic clinical guidelines palliative       \\
6        & 0.07577          & data imaging model method analysis sensitivity based accuracy volume study values methods detection performance results diagnostic mri compared images test                      \\
7        & 0.07482          & case tumors pancreatic thyroid tumor cases diagnosis carcinoma patient lesions malignant adenocarcinoma report rare imaging biopsy benign metastatic petct primary               \\
8        & 0.07106          & expression mir cancer gene tissues cell genes analysis tumor lung crc protein study normal significantly tissue mirnas correlated prognosis rna                                  \\
9        & 0.06988          & patients treatment therapy months radiation radiotherapy dose chemotherapy median treated brain metastases local cancer lung tumor received control survival followup            \\
10       & 0.06737          & protein signaling dna cell proteins binding role pathway activity cells transcription human cellular regulation function receptor activation gene complex show                   \\
\end{tabular}
\label{table:t2}
\end{table}

\section{Topic Modelling on PubMed}
\label{S:3}

In the previous sections, we utilized numerous visualization techniques to visually explore and get a general overview of the PubMed corpora. In this section, we use the MALLET library to perform topic modeling to extract knowledge from the biomedical database by identifying and clustering the major topics inherent in the database.

MALLET is a Java-based package that includes sophisticated tools and a wide variety of algorithms for performing statistical natural language processing tasks like document classification, information extraction, topic modeling, etc., for analyzing large collections of dark data and extracting information out of it. It is co-written by Andrew McCallum and his group at the University of Massachusetts Amherst, as well as contributions from Fernando Pereira, Ryan McDonald, and others at the University of Pennsylvania. The topic modeling toolkit in MALLET contains several efficient, sampling-based implementations of Latent Dirichlet Allocation (LDA), Pachinko Allocation, and Hierarchical LDA, etc.

We use Latent Dirichlet Allocation from the topic modeling toolkit to map terms and words present in the database into a low dimensional continuous space by exploiting the word collocation patterns. We then use Kmeans, a famous clustering algorithm to group words into semantically related clusters according to their cosine similarity measure. Typically, only a small number of topics are present in each document, and only a small number of words have a high probability in each topic. So, we represent these topics along with top-n most frequent terms in each topic using a Topic Cloud.

A Topic Cloud is a pie chart visualization of inherent topics inside a database which consists of a number of topic slices, where each slice contains the most important words in that topic. The relative prominence of words in a topic is made explicit by scaling their sizes in proportion to their confidence score as computed by LDA. A topic cloud is like a word cloud giving the frequency of words or phrases, but the major difference that a topic cloud offers as compared to word cloud is the semantic grouping of words under similar topics, hence capturing the contextual relationship between terms and providing more significant insights into the text data with refined granularity.

We analyse top \textit{n} most significant topics inherent in PubMed where \textit{n} specifies the number of topics considered. Figure~\ref{fig:f5} shows the topic clouds for \textit{n=5} and \textit{n=7} topics respectively. Both the topic clouds elegantly summarize the contents of PubMed documents, with each slice depicting the most important words in that topic. Words like \textit{patients, cancer, disease, treatment} etc. are grouped under one topic whereas words such as \textit{data, study, research} depict another topic. On comparison of both the topic clouds, we find that the topics produced for \textit{n=7} criterion are more homogenous, with uniform proportions, while the topics generated for \textit{n=5} criterion are more disproportionate; indicating \textit{n=7} to be a coherently less noisy criterion as compared to the \textit{n=5} criterion. In addition to topic clouds representing top five and seven most significant topic, we also compute top 10 topics inherent in PubMed as reported in Table~\ref{table:t2} along with proportion for each topic representing the confidence score as computed by LDA. The topic with most significant proportions majorly focuses upon recent research and development strategies such as \textit{molecular and therapeutic research} primarily concerned with \textit{Cancer}, indicating substantial research work related to Cancer.

\section{Bibliometric Citation Analysis using Network Graphs}
\label{S:4}

Network visualization also called Network graphs are often used to visualize complex and convoluted relations between an enormous amount of entities. They represent information in a hierarchically structured manner through an interconnected network of entities highlighting the correlation between them. At its most basic level, a network graph consists of nodes and edges. Nodes represent the entities, and edges represent the relationship between those entities. Edges in the graph can be directed or undirected. Directed edges indicate the flow of information from one node to another. Undirected edges, on the other hand, show the presence of a bidirectional relationship between the two nodes.

\begin{figure}[!p]
\centering
\subfloat[ ][ ]{\includegraphics[scale=0.36]{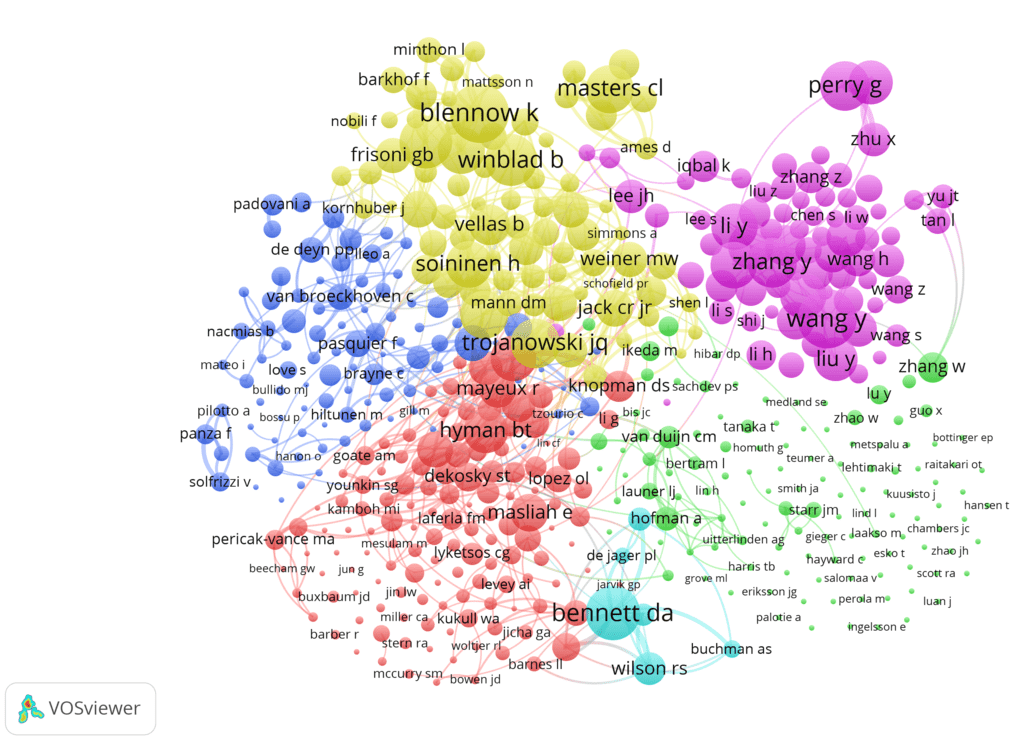}} \\
\subfloat[ ][ ]{\includegraphics[scale=0.36]{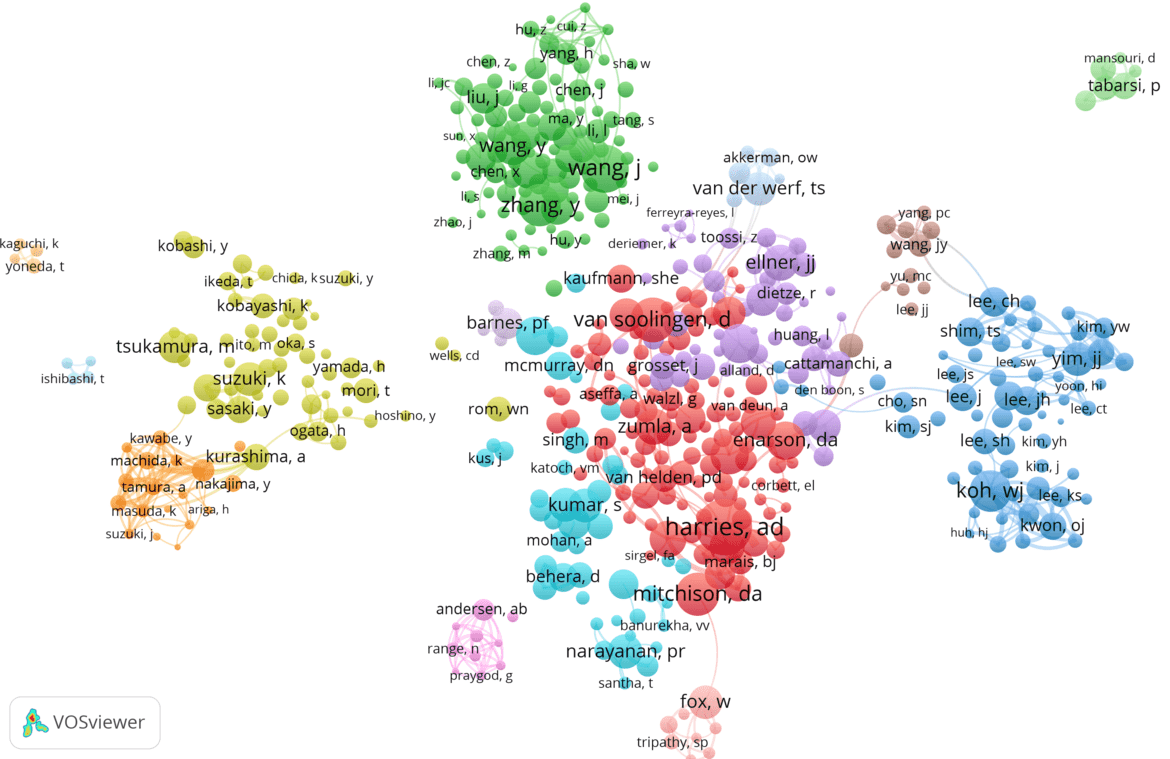}}
\caption{Co-Authorships Networks for (a) Alzheimer’s Disease and (b) Tuberculosis generated using VOSviewer, network and citation viewer.}
\label{fig:f6}
\end{figure}

One of its variants, Citation networks has been used extensively in the field of bibliometrics \citep{Author11year11}. Citation networks proffer quick summarization and visualization of the structure inherent to a set of publications. The resulting visualizations provide insights not only into the present state of scientific research but in identifying potential future research directions and collaboration opportunities. Bibliometric or citation networks can be classified into \textit{direct citation networks, co-citation networks,} and \textit{bibliographic coupling relations}. Direct citation networks, also known as cross citation networks represent research documents citing each other directly as nodes in the network. These networks only offer a direct indication of relatedness between the entities. These are usually very sparse networks and hence relatively uncommon in research settings.

The second variant of citation networks, i.e., co-citation networks represents co-cited research documents (i.e., a pair of documents that are cited by some other group of common documents) as network entities. The higher the number of research documents citing the two documents, the stronger is the relatedness between them. \citep{Author12year12} used these co-citation networks to study the researchers in the field of information science. And in the final variant, bibliographic coupling, two documents are said to be coupled if both cite a common research document \citep{Author13year13}. In other words, the more common the references two documents have, the stronger is the coupling relation between them.

 We use VOSviewer, “an open source software tool used for constructing and visualizing bibliometric networks”, \citep{Author2year2} to construct networks of scientific publications and journals. Each visualization map consists of a network of objects of interests, also known as entities. Entities can be research documents, authors, or keywords which are interconnected with other entities through edges representing citation (co-citation and bibliographic coupling), co-authorship, or co-occurrence links. Each link has a strength associated with it, represented by a positive numerical value. The higher this value, the stronger is the link between the connected entities. Furthermore, each entity is grouped into a non-overlapping and exhaustive cluster. Entities have various attributes associated with them for instance, the weight attribute of the entity or the distance between two entities. The weight of an entity indicates the importance of that entity in the network. An entity with a higher weight is regarded as more important than an entity with a lower weight and hence shown more prominently. The distance between any two entities indicates the strength of the relationship between the corresponding entities. The closer the entities are to each other, the stronger they are correlated with each other.

We create two types of visualizations, \textit{Co-Authorship} and \textit{Co-occurrence Word} Networks. The Co-Authorship networks link the authors of various biomedical research publications in Pubmed based upon the number of publications they have co-authored. These networks help in obtaining significant insights on possible communities and group of researchers who are involved in contributing in their field and may prove to be of significant help to researchers of the same or even different field which are closely related to closely focus on the works of major contributors and head their research forward. The Co-occurrence word networks on the other hand link keywords and term phrases which co-occur together. These networks reveal the semantic correlation among different terms along with major terms highlighted within each cluster. Various useful insights for example in case of PubMed names of major medicines used to cure a disease, or side-effects of treatment, or the possible age groups or gender targeted because of disease can be inferred from these networks with ease.

\begin{figure}[!p]
\centering
\subfloat[ ][ ]{\includegraphics[scale=0.37]{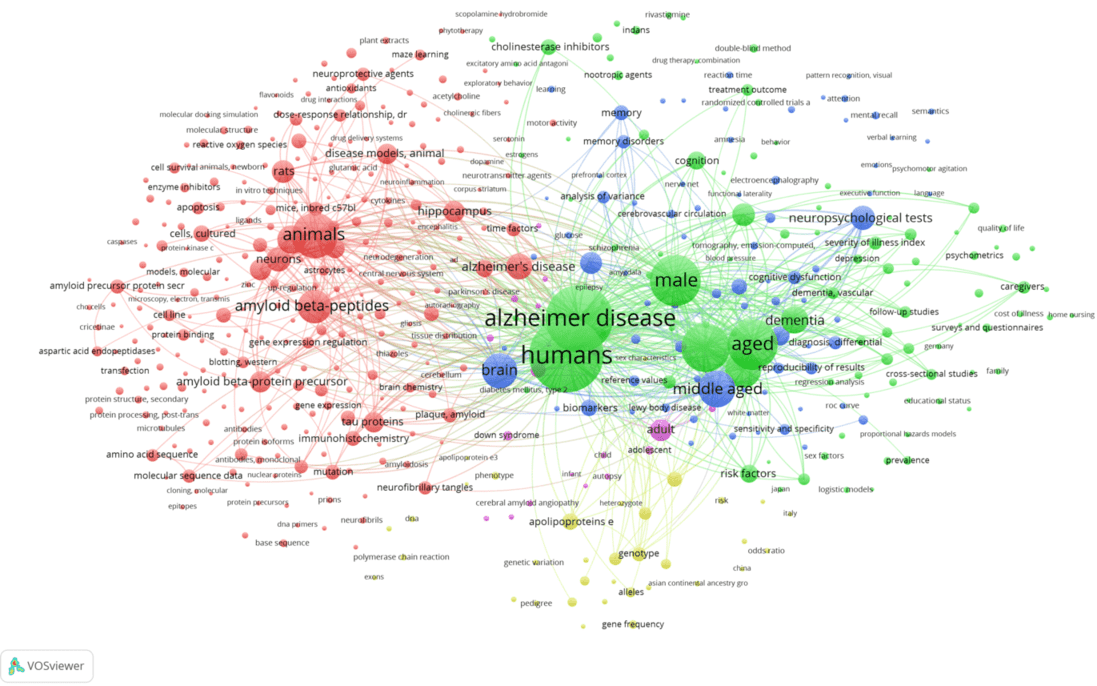}} \\
\subfloat[ ][ ]{\includegraphics[scale=0.37]{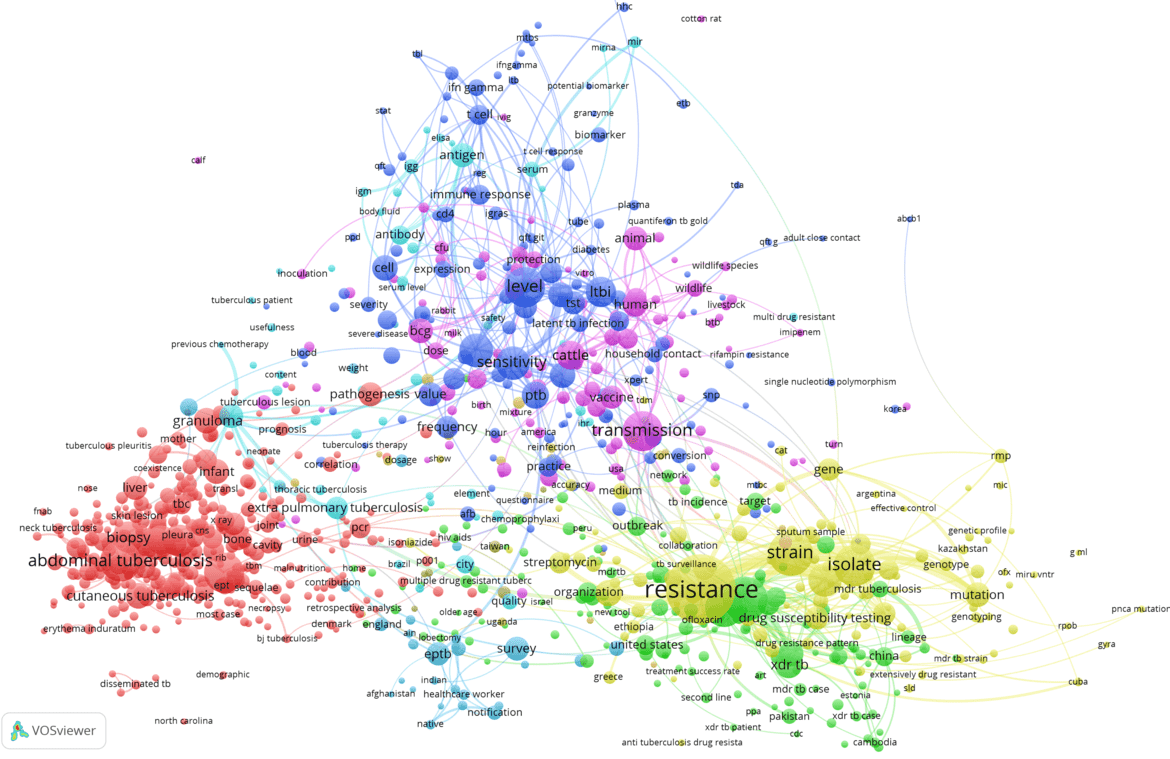}}
\caption{Co-Occurrence Word Networks for (a) Alzheimer’s Disease and (b) Tuberculosis generated using VOSviewer, network and citation viewer.}
\label{fig:f7}
\end{figure}

To create these networks, we query the PubMed database on two different topics, \textit{Alzheimer's Disease} and \textit{Tuberculosis} respectively. We obtain the resulting abstracts for each topic from the PubMed site in MEDLINE format. While creating these networks, some parameters need to be selected. Like for the co-authorship networks, we choose the parameter of \textit{full counting}, i.e., each link contributes equally and Authors as the unit of analysis. We choose the minimum number of documents an author must have published to be ten as the threshold to limit our network to mentions of authors who have contributed at least ten documents related to the topic for which the network map is created. Finally, from the authors shortlisted we select top 500 authors to be visualized in our network based upon their total link strength which indicates the total strength of the co-authorship links of a given researcher with other researchers. Similarly, for the Co-occurrence based Word Networks, we first select the option to ignore copyright statement to get rid of the unwarranted text, we extract text from both title and abstract fields of the MEDLINE document, and we select the option of full counting, to count all the occurrences of a term in a document. We filter the less significant terms from more significant terms by setting the minimum number of occurrences of a term barrier to 8. For each of the filtered term, VOSviewer calculates a relevance score, which represents the specificity of a term towards the topics covered by the text data. We select the top 80\% most relevant terms depending upon the relevance score metric to be displayed in our network. We select minimum cluster size to be 2 and \textit{Association Strength} to be the normalization method for the layout algorithm for both visualizations discussed above.


Figure~\ref{fig:f6}(a) shows the Co-Authorship networks for PubMed documents related to the topic of Alzheimer’s Disease. From the figure, we can observe that some authors like \textit{Bennett Da, Blennow K, Perry G, Zhang Y, Wang Y} have bigger node sizes as compared to other authors thereby indicating a higher proportion of work contributed by these authors in the queried field of work. From the same figure, we can also observe the potential clusters of authors depending upon the papers they have co-authored and the areas of their study. Authors in the purple and yellow cluster appear to work only with authors within their clusters while the work of authors of red, blue and cyan clusters are uniformly interspersed between different clusters. Also, from the distances between two authors in the visualization and the thickness of links connecting them, the relatedness of the authors can be inferred. In general, the closer two authors are located to each other or thicker the link connecting them is, the stronger their relatedness.

Similarly, Figure~\ref{fig:f6}(b) shows the Co-Authorship network for PubMed documents related to the topic of \textit{Tuberculosis}. From the figure, prominent authors can be observed based on their node sizes like \textit{Harries AD, Van Soolingen, Wang J, Narayanan PR,} etc. Various clusters can also be observed shown in distinct colors with the authors in the red cluster can be seen to be widely connected with authors present in other clusters, thereby indicating major source of work done by these authors related to tuberculosis.

In addition to the co-authorship networks, we also show the Co-Occurrence word networks in Figure~\ref{fig:f7}. Various useful insights can be gathered from these networks. Firstly, from the co-occurrence word network shown in Figure~\ref{fig:f7}(a) we can infer that \textit{Alzheimer’s disease} is related to the brain due to presence cooccurring terms such as \textit{brain, memory, cognition,} etc. We can also infer that \textit{males} are more vulnerable to Alzheimer disease as compared to \textit{females}. Potential age group suffering from Alzheimer disease can also be identified as the \textit{middle age} to \textit{old age} group. Various side effects of Alzheimer disease can also be found such as \textit{memory disorders, dementia, depression} etc. 

Similarly, from the co-occurrence word network for \textit{Tuberculosis} shown in Figure~\ref{fig:f7}(b) many important terms such as certain types of tuberculosis, like \textit{abdominal tuberculosis, pulmonary tuberculosis, neck tuberculosis,} etc can be identified. The network also lists the names of certain \textit{vaccines} and \textit{resistance techniques} related to tuberculosis. Finally, on studying the network in depth, names of certain places like \textit{India, North Carolina, England,} etc can be found in association with terms like \textit{healthcare workers, survey, treatment success rate,} etc indicating that these places are playing a major role in spreading information and public awareness related to disease and are providing proper treatment to people affected by tuberculosis.

\section{Conclusion}
\label{S:5}

The main motivation behind the present work was to diminish the limitation of human cognition and perception, in handling and examining enormous amounts of information. The present analytical work endeavored to exploit various data visualization tools and natural language processing techniques to proffer a potential solution to overcome the problem of analyzing overwhelming amounts of information. In this empirical research, we utilized the rich corpus of the PubMed to visually explore and analyze lexical and textual biomedical dark data to mine knowledge out of it. We employed various text summarization and visualization techniques like computation of raw term and document frequencies, word clouds, $TF\times IDF$ scores, DocuBurst, etc., to get a general overview of the PubMed database. We then utilized the MALLET library to perform topic modeling to extract knowledge from the biomedical database and visualized the inherent major topics inherent in PubMed using Topic Clouds. Finally, we used network and citation visualization techniques, to construct bibliometric networks, i.e., Co-Authorship and Co-Occurrence word networks for studying relationships between different entities like scientific documents and journals, researchers, and, keywords and terms. All of the techniques that were employed to explore visually and mine knowledge from dark data, i.e., PubMed proved to be of great help in allowing quick comprehension of information, the discovery of emerging trends, and identification of relationships and patterns within the database.






\bibliographystyle{elsarticle-num-names}
\bibliography{main.bib}







\end{document}